\documentclass[aps,prd,twocolumn,showpacs,superscriptaddress,amsmath,graphicx]{revtex4}

\usepackage{graphicx}

\begin{document}

\title{Note on helicity amplitudes in $D \to V$ semileptonic decays}

\author{Svjetlana Fajfer}
\email[Electronic address:]{svjetlana.fajfer@ijs.si}
\affiliation{J. Stefan Institute, Jamova 39, P. O. Box 3000, 1001 Ljubljana, Slovenia}
\affiliation{Department of Physics, University of Ljubljana, Jadranska 19, 1000 Ljubljana, Slovenia}

\author{Jernej Kamenik}
\email[Electronic address:]{jernej.kamenik@ijs.si}
\affiliation{J. Stefan Institute, Jamova 39, P. O. Box 3000, 1001 Ljubljana, Slovenia}

\date{\today}

\begin{abstract}
Motivated by the recent extraction of the helicity amplitudes for the  
$D^+ \to \bar K^{*0} \mu \nu_{\mu}$  decay, done by the FOCUS 
collaboration,  
we determine helicity amplitudes for the $D^+ \to \bar K^{*0} l \nu_l$, $D^+ \to \rho^0 l \nu_l$ 
and $D^+_s \to \phi l \nu_l$ semileptonic decays 
using the knowledge of the 
relevant form factors. The vector and axial form 
factors for $D \to V l \nu_l$ 
decays are parameterized by including contributions of charm meson resonances and using the 
HQET and SCET limits. In the case that the vector form factor receives 
contributions from two poles while axial form factors are dominated by a single pole for 
$D^+ \to \bar K^{*0} l \nu_l$, 
we obtain better  agreement with the 
experimental result 
 then when all of them are 
dominated by single poles.  

\end{abstract}


\maketitle

Recently the FOCUS~\cite{Link:2005dp} collaboration has presented first 
non-parametric determination of helicity amplitudes in the semileptonic decay 
$D^+\to\bar K ^{*0} \mu^+ \nu$. This measurement allows for more detailed analysis of the $D\to V$ form factors, especially it enables the studying of the shapes of the form factors. 

\par

We have recently proposed a generalization of the Be\'cirevi\'c-Kaidalov (BK) form factor 
parameterization~\cite{Becirevic:1999kt} for the semileptonic $H\to V$ form 
factors based on HQET and SCET scaling predictions~\cite{Fajfer:2005ug}. Furthermore we have
calculated the $D\to P$ and $D\to V$ form factors shapes within a model which combines
properties of the heavy meson chiral Lagrangian by  taking into account known and predicted charm resonances and utilizing the general form factor parameterizations~\cite{Fajfer:2004mv,Fajfer:2005ug}. 

\par

In this note we determine helicity amplitudes for the  $D\to V$ semileptonic 
decays and 
compare our model predictions for the shapes of the form factors 
with the new experimental results coming from FOCUS 
for the $D^+\to\bar K ^{*0} \mu^+ \nu$ decay. 

\par

The standard decomposition of the current matrix elements 
relevant to semileptonic decays between a heavy pseudoscalar meson state 
$|H(p_H)\rangle$ with momentum $p_H$ and a light vector meson 
$|V(p_V,\epsilon_V)\rangle$ with momentum 
$p_V$ and polarization vector $\epsilon_V$ is in terms of four 
form factors $V$, $A_0$, $A_1$ and $A_2$, functions of the exchanged momentum 
squared $q^2 = (p_H-p_V)^2$~\cite{Wirbel:1985ji}. Here $V$ denotes the vector form factor and is expected to be dominated by vector meson resonance exchange, the axial $A_1$ 
and $A_2$ form factors are expected to be dominated by axial resonances, 
while $A_0$ denotes the pseudoscalar form factor and is expected to be dominated by pseudoscalar meson resonance exchange~\cite{Wirbel:1985ji}.
In order that the matrix elements are finite at $q^2=0$, the form factors must also satisfy the well known relation $A_0(0)+A_1(0)(m_H+m_V)/2m_V-A_2(0)(m_H-m_V)/2m_V=0$.
\par
Next we follow the analysis of Ref.~\cite{Becirevic:1999kt}, where the $F_+$
form factor in $H\to P$ transitions is given as a sum of two pole 
contributions, while the
$F_0$ form factor is written as a single pole, based on form factor 
dispersion properties as well as known HQET~\cite{Isgur:1990kf} and 
SCET~\cite{Charles:1998dr,Beneke:2000wa,Ebert:2001pc} scaling limits near zero and maximum recoil 
momentum respectively. Utilizing the same approach we have proposed a general 
parametrization of the heavy to light vector form factors, which also takes 
into account all the known scaling and resonance 
properties of the form factors. The details of the analysis are outlined 
in Ref.~\cite{Fajfer:2005ug} and we only give the results for the derived 
form factor parameterizations:
\begin{eqnarray}
V(q^2) &=& \frac{c'_H (1-a)}{(1-x)(1-a x)},\nonumber\\
A_1(q^2) &=&  \xi \frac{c'_H(1-a)}{1-b' x},\nonumber\\
A_0(q^2) &=& \frac{c''_H (1-a')}{(1-y)(1-a' y)},\nonumber\\
A_2(q^2) &=& \frac{c'''_H}{(1-b' x)(1-b'' x)},\nonumber\\
\end{eqnarray}
where $c'''_H = [(m_H+m_V) \xi c'_H (1-a) + 2 m_V c''_H (1-a')]/(m_H-m_V)$ is 
fixed by the relation between the form factors at $q^2=0$ while 
$\xi=m_H^2/(m_H+m_V)^2$ is the proportionality factor between $A_1$ and $V$ 
from the SCET relation. Variables $x=q^2/m_{H^*}^2$ and $y = q^2/m_H^2$ ensure, that 
the $V$ and $A_0$ form factors are 
dominated by the physical $1^-$ and $0^-$ resonance poles, while $a$ and $a'$ 
measure 
the contributions of higher states, parameterized by additional 
effective poles. On the other hand $b'$ in $A_1$ and $A_2$ measures the 
contribution of resonant states with spin-parity assignment $1^+$ which are 
parameterized by the effective pole at $m_{H'^*_{\mathrm{eff}}}^2=m_{H^*}^2/b'$
 while the scaling properties and form factor relations require an additional
 effective pole for the $A_2$ form factor. At the end we have parameterized the
 four $H\to V$ vector form factors in 
terms of the six parameters $c'_H$, $a$, $a'$, $b'$, $c''_H$ and 
$b''$.

\par

We determine the above parameters via heavy meson chiral theory (HM$\chi$T) calculation of the form factors near $q^2_{\mathrm{max}}=(m_H-m_V)^2$. We use the leading 
order heavy meson 
chiral  Lagrangian in which we include additional charm meson states.
The details of this framework are given in \cite{Fajfer:2005ug}  and 
\cite{Fajfer:2004mv}. 
We first calculate values of the form factors in the small recoil region. 
The presence of charm meson resonances in our Lagrangian affects the values of
the form factors at $q^2_{\mathrm{max}}$ and induces saturation of the second 
poles in the parameterizations of the $F_+(q^2)$, $V(q^2)$ and $A_0(q^2)$ form 
factors by the next radial excitations of $D_{(s)}^*$ and $D_{(s)}$ mesons 
respectively. 
Using HQET parameterization of the current matrix 
elements~\cite{Fajfer:2005ug}, which is especially suitable for HM$\chi$T 
calculations of the form factors near zero recoil, we are able to extract 
consistently the contributions of individual resonances from our Lagrangian 
to the various $D\to V$ form factors. 
We use physical pole masses of excited state charmed mesons in the 
extrapolation, giving for 
the pole parameters $a=m_{H^{*}}^2/m_{H'^{*}}^2$, $a'=m_{H}^2/m_{H'}^2$ and 
$b'=m_{H^*}^{2}/m_{H_{A}}^2$.
Although in the general parameterization of the form factors the extra poles 
in 
$V$ and $A_{0,1,2}$ parameterize all the neglected higher resonances beyond 
the ground state heavy meson spin doublets $(0^-,1^-)$, we are
saturating those by a single nearest resonance.
The single pole $q^2$ behavior of the $A_1(q^2)$ form factor is explained 
by the presence of a single $1^+$ state relevant to each decay, while in 
$A_2(q^2)$ in addition to these states one might also account for their next 
radial excitations. However, due to the lack of data on their presence we 
assume their masses being much higher than the first $1^+$ states and we 
neglect their effects, setting effectively $b''=0$.

\par

The values of the unknown HM$\chi$T parameters appearing in $D \to V l \nu_l$ 
decay amplitudes~\cite{Fajfer:2005ug} 
are determined by fitting the model predictions to known experimental
values of branching ratios and partial decay width ratios.

\par 

In order to compare our model predictions with recent experimental analysis performed by 
FOCUS collaboration, following ~\cite{Ball:1991bs} 
 we introduce helicity amplitudes $H_{+,-,0}$: 
\begin{eqnarray}
H_{\pm}(y) &=& + (m_H+m_V) A_1(m_H^2 y) \mp \frac{2 m_H |\vec p_V(y)|}{m_H+m_V} V(m_H^2 y)\nonumber\\*
H_0(y) &=& + \frac{m_H+m_V}{2 m_H m_V \sqrt y} [ m_H^2 (1-y) -m_V^2] A_1(m_H^2 y)\nonumber\\*
	&& - \frac{2 m_H |\vec p_V(y)|}{m_V(m_H+m_V) \sqrt y } A_2(m_H^2 y)
\end{eqnarray} 
where $y=q^2/m_H^2$ and the three-momentum of the light vector meson is given by: 
\begin{equation}
|\vec p_V (y)|^2 = \frac{[m_H^2 (1-y) + m_V^2]^2}{4 m_H^2} -m_V^2.
\end{equation}

\par

\begin{figure}[!h]
\scalebox{0.9}{\includegraphics{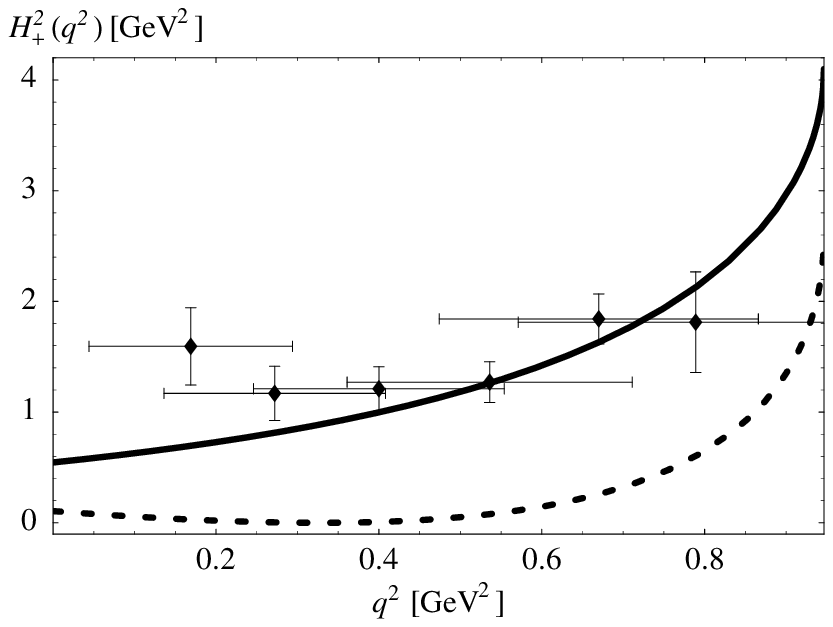}}
\caption{\label{1} Our model predictions (double pole in solid line and single pole in dashed line) for the $q^2$ dependence of the 
helicity amplitude $H_+^2(q^2)$ in comparison with scaled FOCUS data on $D^+ \to \bar K^{*0}$ semileptonic decay.}
\end{figure}

\begin{figure}[!h]
\scalebox{0.9}{\includegraphics{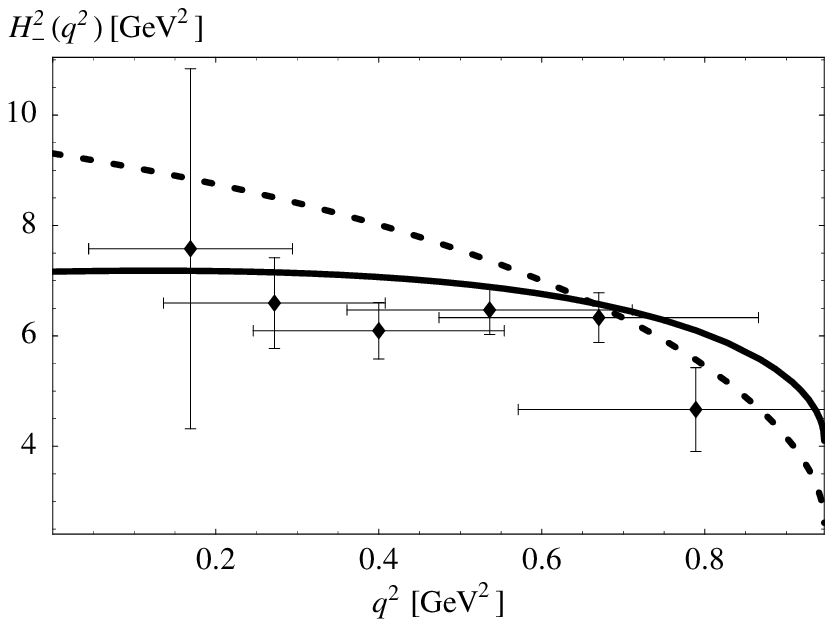}}
\caption{\label{2} Our model predictions (double pole in solid line and single pole in dashed line) for the $q^2$ dependence of the
 helicity amplitude $H_-^2(q^2)$ in comparison with scaled FOCUS data on $D^+ \to \bar K^{*0}$ semileptonic decay.}
\end{figure}

\begin{figure}[!h]
\scalebox{0.9}{\includegraphics{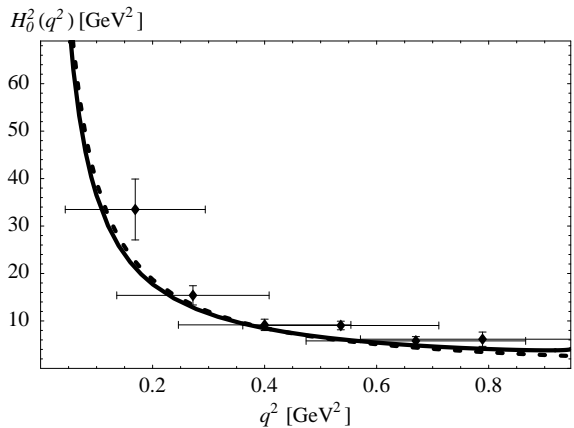}}
\caption{\label{3} Our model predictions (double pole in solid line and single pole in dashed line) for the $q^2$ dependence of the 
helicity amplitude $H_0^2(q^2)$ in comparison with scaled FOCUS data on $D^+ \to \bar K^{*0}$ semileptonic decay.}
\end{figure}

Because of the arbitrary normalization of the form factors 
in~\cite{Link:2005dp}, we fit our model predictions for a common overall 
scale in order to compare the results. We plot the $q^2$ dependence of the 
predicted helicity amplitudes and compare them with the experimental 
results of FOCUS, scaled by an overall factor determined by the least 
square fit of our model predictions, on FIGs.~\ref{1},~\ref{2} and~\ref{3}. 
The scale factor is common to all form factors. 

\begin{figure}[!h]
\scalebox{0.9}{\includegraphics{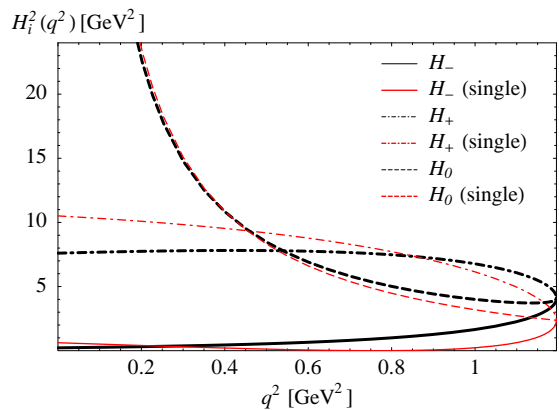}}
\caption{\label{4} Our model predictions for the $q^2$ dependence of the helicity amplitudes $H_i^2(q^2)$ for the $D^+\to \rho^0$ semileptonic decay. Double pole predictions are rendered in thick (black) lines while single pole predictions are rendered in thin (red) lines: $H_-$ (solid lines), $H_0$ (dashed lines) and $H_+$ (dot-dashed lines).}
\end{figure}

\begin{figure}[!h]
\scalebox{0.9}{\includegraphics{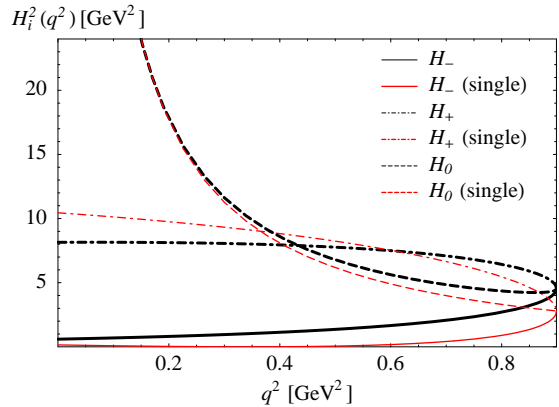}}
\caption{\label{5} Our model predictions for the $q^2$ dependence of the helicity amplitudes $H_i^2(q^2)$ for the $D_s^+\to \phi$ semileptonic decay. Double pole predictions are rendered in thick (black) lines while single pole predictions are rendered in thin (red) lines: $H_-$ (solid lines), $H_0$ (dashed lines) and $H_+$ (dot-dashed lines).}
\end{figure}

In addition to the two pole contributions we calculate helicity amplitudes 
in the case when all the form factors exhibit single pole behavior. 
Putting contributions of higher charm resonances to be zero we fit the remaining model parameters to existing branching ratios and partial decay ratios. We obtain the values for the following parameter combinations as explained in~\cite{Fajfer:2005ug}:
\begin{eqnarray}
\tilde\alpha\tilde\mu &=& 0 \nonumber\\*
\alpha'\zeta &=& -0.180~\mathrm{GeV}^{3/2} \nonumber\\*
\alpha'\mu &=& -0.00273~\mathrm{GeV}^{1/2} \nonumber\\*
\alpha_1 &=& -0.203~\mathrm{GeV}^{1/2}
\end{eqnarray}
As shown on FIGs.~\ref{1} and~\ref{2} the experimental data for $H_{\pm}$ do not favor such a parametrization, while in the case of $H_0$ helicity amplitude there is almost no difference since the $H_0$ helicity amplitude is defined via the $A_{1,2}$ form factors, which are in our approach both effectively dominated by a single pole. 
The agreement between the FOCUS results and our model predictions for the 
$q^2$ dependence of the helicity amplitudes is good, although as noted already 
in~\cite{Link:2005dp}, the uncertainties of the data points are still rather large. 
On FIGs.~\ref{4} and~\ref{5} we present helicity amplitudes for the $D^+ \to \rho^0 l \nu_l$ and 
$D^+_s \to \phi  l \nu_l$ decays. Both decay modes are most promissing for the future experimental studies. 
We make predictions for the shapes of helicity amplitudes for both cases: where two poles contribute to the vector form factor and a single pole to the axial form factors, and the second case where all form factors exhibit single pole behavior.

\par

In principle one can apply the above procedure to the $B \to \rho l \nu_l$ 
semileptonic decays. However, due to the much broader leptons invariant mass dependence in 
this case, our procedure is much more sensitive to the values of the form factors at $q^2 \approx  0$. 
In addition, the semileptonic decay rates in our model fit are numerically 
dominated by the longitudinal helicity amplitude $H_0$ which has a broad 
$1/\sqrt{q^2}$ pole~\footnote{Naive HQET scaling predicts that the $H_-$ 
helicity amplitude, which scales as $\sqrt{m_H}$ should dominate the decay 
rate.}. This is true especially for $D\to V$ but to minor extent also for 
$B\to V$ transitions. Since our model parameters are determined at 
$q^2_{\mathrm{max}}$, this gives a poor handle on the dominating effects in 
the overall decay rate. Thus, accurate determination of the magnitude and 
shape of the $H_0$ helicity amplitude near $q^2=0$ would contribute much to 
clarifying this issue.

\par

We can summarize:
we have investigated the predictions of the general $H\to V$ form factor 
parametrization combined with HM$\chi$T calculation for the $D^+ \to \bar K^{*0}$ 
semileptonic helicity amplitudes, recently determined by the FOCUS collaboration. 
In addition we have determined the helicity amplitudes for the $D^+ \to \rho^0 l \nu_l$ and $D_s^+ \to \phi l \nu_l$ decays. 
In all three cases that we have considered we used two approaches: one with a 
two poles shape for the vector form factor and single pole for the axial form factors, and secondly the usually assumed single pole behavior of all three relevant form factors. 
Our study indicates that the two pole shape for the $V(q^2)$ form factor in $D^+ \to \bar K^{*0}$ transition is favored over the single pole shape, when compared to the FOCUS result.



\begin{acknowledgments}
We are thankful to D. Kim and J. Wiss from the FOCUS collaboration for sending us their data and for helping us understand it.
This work is supported in part by the Ministry of Higher Education, Science and Technology of the Republic of Slovenia.
\end{acknowledgments}

\bibliography{article}

\end{document}